\documentclass[onecolumn,final,letterpaper,12pt,journal]{IEEEtran} 
\pdfoutput = 1
\usepackage{amsmath,amssymb,amsfonts,mathtools}
\usepackage{graphicx}
\usepackage{cite}
\usepackage{color}
\usepackage{enumerate}
\usepackage{makecell}

\usepackage[mathscr]{euscript}
\usepackage{stmaryrd}
\usepackage{multirow}
\usepackage{bm,upgreek}
\usepackage{framed}
\usepackage{diagbox}

\usepackage{algorithm,algorithmic}
\usepackage{setspace}

\newcommand{\diag}{\mathrm{diag}}
\newcommand{\blkdiag}{\mathrm{block\textrm{-}diag}}

\newcommand{\E}{\mathcal{E}}
\newcommand{\diff}{\mathrm{d}}
\newcommand{\Trans}{\mathrm{T}}

\def\MatrixFont{}
\def\VectorFont{\bf}

\newcommand{\mA}{{\MatrixFont A}}
\newcommand{\mB}{{\MatrixFont B}}

\newcommand{\mF}{{\MatrixFont F}}
\newcommand{\mG}{{\MatrixFont G}}
\newcommand{\mH}{{\MatrixFont H}}
\newcommand{\mI}{{\MatrixFont I}}

\newcommand{\mL}{{\MatrixFont L}}

\newcommand{\mP}{{\MatrixFont P}}
\newcommand{\mQ}{{\MatrixFont Q}}
\newcommand{\mR}{{\MatrixFont R}}

\newcommand{\mV}{{\MatrixFont V}}
\newcommand{\mW}{{\MatrixFont W}}
\newcommand{\mX}{{\MatrixFont X}}
\newcommand{\mY}{{\MatrixFont Y}}

\newcommand{\va}{{\VectorFont a}}

\newcommand{\vv}{{\VectorFont v}}
\newcommand{\vw}{{\VectorFont w}}
\newcommand{\vx}{{\VectorFont x}}
\newcommand{\vy}{{\VectorFont y}}
\newcommand{\vz}{{\VectorFont z}}

\newcommand{\vones}{{\VectorFont 1}}
\newcommand{\mf}{f}
\newcommand{\mh}{h}
\newcommand{\mk}{k}
\newcommand{\mg}{g}

\newcommand{\vnu}{\bm \nu}
\newcommand{\vxi}{{\bm \xi}}
\newcommand{\vbeta}{{\bm \beta}}
\newcommand{\valpha}{{\bm \alpha}}

\newcommand{\vGamma}{{\Gamma}}

\newcommand{\vLambda}{{\Lambda}}
\newcommand{\sizehat}{\widehat}
\newcommand{\sizetilde}{\widetilde}
\newcommand{\gram}{K}

\def\H{\mathcal{H}}

\def\Y{\mathbb{Y}}

\def\HS{\mathrm{HS}}
\def\nP{m}
\def\dx{n_x}
\def\dy{n_{y_i}}
\def\dY{n_y}
\def\R{\mathbb{R}}
\def\N{\mathcal N}
\def\Na{\mathbb N}
\def\nS{n}
\def\p{p}
\def\G{\mathcal G}
\def\V{\mathcal V}
\def\F{\mathcal F}
\def\G{\mathcal G}

\newtheorem{theorem}{Theorem}
\newtheorem{lemma}{Lemma}
\newtheorem{proposition}{Proposition}
\newtheorem{remark}{Remark}

\newtheorem{assumption}{Assumption}
\newtheorem{corollary}{Corollary}

\renewcommand{\IEEEQED}{\IEEEQEDopen}
\IEEEoverridecommandlockouts
\allowdisplaybreaks[3]
\begin{document}
\title{Consensus-Based Distributed Nonlinear Filtering with Kernel Mean Embedding
\thanks{
	The work was supported by the National Key R\&D Program of China under Grant 2018YFA0703800, National Natural Science Foundation of China under Grants 62203045 and T2293770 and 62025306, CAS Project for Young Scientists in Basic
	Research under Grant YSBR-008. \emph{(Corresponding author: Ji-Feng Zhang.)}}
\thanks{Liping Guo, Yanlong Zhao and Ji-Feng Zhang are with the Key Laboratory of Systems and Control, Institute of Systems Science, Academy of Mathematics and Systems Science, Chinese Academy of Sciences, Beijing 100190, China. Yanlong Zhao and Ji-Feng Zhang are also with the School of Mathematical Sciences, University of Chinese Academy of Sciences, Beijing 100049, China (e-mail: lipguo@outlook.com; ylzhao@amss.ac.cn; jif@iss.ac.cn).}
\thanks{Jimin Wang is with the School of Automation and Electrical Engineering, University of Science and Technology Beijing, Beijing 100083, and also with the Key Laboratory of Knowledge Automation for Industrial Processes, Ministry of Education, Beijing
	100083, China (e-mail: jimin.wang@amss.ac.cn).}}
\author{Liping Guo, Jimin Wang, \IEEEmembership{Member,~IEEE}, Yanlong Zhao,~\IEEEmembership{Senior Member,~IEEE},
and Ji-Feng Zhang,~\IEEEmembership{Fellow,~IEEE}}

\maketitle
\IEEEpeerreviewmaketitle
\thispagestyle{plain}

\begin{abstract}
	This paper proposes a consensus-based distributed nonlinear filter with kernel mean embedding (KME).
	This fills with gap of posterior density approximation with KME for distributed nonlinear dynamic systems. To approximate the posterior density, the system state is embedded into a higher-dimensional reproducing kernel Hilbert space (RKHS), and then the nonlinear measurement function is linearly converted. As a result, an update rule of KME of posterior distribution is established in the RKHS.
	To show the proposed distributed filter being capable of achieving the centralized estimation accuracy, a centralized filter, serving as an extension of the standard Kalman filter in the state space to the RKHS, is developed first. Benefited from the KME, the proposed distributed filter converges to the centralized one while maintaining the distributed pattern. 
	Two examples are introduced to demonstrate the effectiveness of the developed filters in target tracking scenarios including nearly constantly moving target and turning target, respectively, with bearing-only, range and bearing measurements.
\end{abstract}


\section{Introduction}

Nonlinear filtering problem is critically important and of great interest in many research fields, such as target tracking \cite{Ristic-Arulampalam-GordonGordon2004,JIA2013High, Lan2015Nonlinear, Lan-Li-2017-Multiple,Guo2022Nonlinear, Liu2023Randomized, Tan2023Nonlinear}, navigation \cite{Kutschireiter2022}, detection \cite{Fisch2022} and medical image processing \cite{Zhu2021Fast}. 
In practice, the measurements are usually collected by some sensors with a network structure \cite{Mahler2014Advances}.
The conventional centralized filtering can achieve a high level of filtering accuracy, but requires each sensor to send their own measurement to a central processing unit in a timely manner. Therefore, it is often burdened with a computationally demanding task and susceptible to processing unit failures.
In contrast, the distributed filtering requires each sensor to implement filtering independently using its own measurement and communicating with its neighbors by graph topology, and thus is generally more practical, robust, resilient and computationally efficient than the centralized one \cite{Hlinka2013Distributed,Xu2017Distributed}.
Consequently, developing effective distributed nonlinear filtering is an attractive topic that warrants attention.


From \cite{Haykin2009Cubature}, the existing nonlinear filtering algorithms focus either on the estimated quantities such as the moments, or on the density.
Based on approximating the first two moments of the posterior density, the distributed Kalman filter (DKF) gaining significant popularity is widely used across various fields \cite{Olfati-Saber2005Consensus,Olfati-Saber2005Distributed,Olfati-Saber2007Distributed,Talebi2019}. By employing the extended Kalman filter, the DKF can be directly applied to distributed nonlinear dynamic systems, but may diverge for highly nonlinear problems \cite{julier2004unscented}. To overcome the limitations of the extended Kalman filter, the distributed unscented Kalman filter \cite{Li2012Distributed}, the distributed cubature Kalman filter \cite{Chen2017}, and their extension \cite{Xiao2022Multisensor} were proposed, respectively. These moment based filters are efficient and convenient to implement, but cannot obtain the global posterior density for further statistical inference.
Based on approximating the posterior density, the particle filter adopts the popular Monte Carlo method to draw a set of particles for the density approximation. With different strategies, the distributed particle filter was developed and widely applied \cite{Gu2008Consensus,Hlinka2011Distributed,Hlinka2014Consensus,Xia2020Direct,Song2021Distributed,Song2021Particle}. 
While the Monte Carlo method is valuable for approximating the posterior density, it has several drawbacks such as heavy computational burden and sensitivity to noise. 
To avoid these drawbacks, the kernel mean embedding (KME) is powerful and widely recognized as an effective density approximation method \cite{Smola2007A,Song2009,sriperumbudur2010hilbert,Song2013,Kanagawa2014Monte,Muandet2017}.
By embedding a probability distribution through kernel function into a higher-dimensional reproducing kernel Hilbert space (RKHS),
the KME represents the posterior density as a point in the RKHS where standard mathematical operations can be performed. Some KME based filters with single sensor using either the kernel Bayes' rule \cite{Fukumizu2013Kernel} or the kernel Kalman rule \cite{Gebhardt2017The} can be found in \cite{Song2009,Kanagawa2014Monte,Sun2023Adaptive}. 
For distributed nonlinear dynamic systems, posterior density approximation with KME is still a gap to be filled.

The merits of the KME mainly lie in the following four aspects:
i) computational efficiency: without repeated sampling and averaging in Monte Carlo method, KME significantly reduces computation time, which is particularly crucial for real-time problems such as filtering;
ii) widespread applications: as a non-parametric method, KME does not require assumptions about the state distribution and can address complicated posterior density often encountered in nonlinear filtering;
iii) dimensionality curse avoided: by using kernel trick, KME implicitly captures complex relationships between prediction and posterior distributions, and thus avoids the curse of dimensionality;
iv) robustness: rather than directly modeling the underlying stochastic process, KME relies on a fixed kernel function and thus can capture the structure of the posterior density even if the input state contains some noise or missing values. 
Overall, the KME is powerful for designing both effective and efficient distributed nonlinear filtering based on posterior density approximation.

In this paper, we aim at developing a consensus-based distributed nonlinear filtering with KME. 
Specifically, we aim at obtaining the embeddings of the prediction and posterior distributions of state, respectively. 
In the prediction step, the challenge lies mainly in selecting the sample points of the prediction distribution. To overcome this challenge, we adjust the positions of the sample points from the previous posterior distribution only while retaining their weights, and keep the predicted sample points for each node consistent for the subsequent consensus.
In the update step, the difficulty in obtaining the posterior distribution and its moments arises from the nonlinearity of the measurement function. By representing a real-valued function as an inner product in the higher-dimensional RKHS, the original nonlinear measurement function defined on the state space is represented as a linear operator defined on the RKHS. Then, an update rule of the KME is established correspondingly
and allows for a direct acquisition of a posterior sample from the KME. As a result, the pre-image problem of obtaining the state estimate in the state space becomes simple. 
For the proposed distributed nonlinear filter, we also expect it to achieve the centralized estimation accuracy. 
To this end, we first provide a centralized filter, which is essentially an extension of the standard Kalman filter in the state space to the RKHS. Then, we prove that the proposed distributed filter converges to the centralized one. Finally, we demonstrate the developed filters through two target tracking examples, compared with an effective DKF.

The contributions of this paper include mainly the following:
\begin{enumerate}
	\item We develop a consensus-based distributed nonlinear filter with KME estimates of posterior distributions, which fills with gap of posterior density estimation with KME for distributed nonlinear dynamic systems. 
	\item We provide a centralized nonlinear filter which extends the standard Kalman filter in the state space to the RKHS and degenerates to the centralized Kalman filter for linear dynamic systems.
	\item We prove that the proposed distributed filter converges to the centralized one, which indicates that the distributed filter brings together the benefits of distributed pattern with the centralized estimation accuracy. Furthermore, we demonstrate the superiority of the developed filters through two different target tracking scenarios with bearing-only, range and bearing measurements.
\end{enumerate}

This paper is organized as follows. In Section \ref{sec:problem formulation}, some preliminaries and problem formulation are given.
In Section \ref{sec:DNF}, the consensus-based distributed nonlinear filter is developed. The centralized nonlinear filter and convergence analysis are provided in Section \ref{sec:optimality}, followed by two examples of target tracking in Section \ref{sec:examples}. Section \ref{sec:conclusion} concludes the paper.

\emph{Notations:}
The scalars, vectors and matrices are denoted by lowercase, bold lowercase and lightface capital letters, respectively. For simplicity and convenience, scalar $0$, zero vector and zero matrix are all denoted by $0$.
$\Na^+$ stands for the set of all positive integers.
The sets of all $p$-dimensional real vectors and all $p \times q$ real matrices are denoted by $\R^{p}$ and $\R^{p \times q}$, respectively.   
For a vector $\va$, $\diag(\va)$ represents a diagonal matrix with principal diagonal consisting of $\va$;
$\| \va \|_q$ denotes $\ell_q$ norm of $\va$; when $q = 2$, it is Euclidean norm abbreviated as
$\| \va \|$.
All vectors in this paper are column vectors.
$\mI$ is an identity matrix with appropriate dimension.
$\blkdiag(\mA_1, \dots, \mA_p)$ represents a block diagonal matrix with main diagonal block consisting of matrices $\mA_i$ ($i = 1, \dots, p$).
The notation ``$\otimes$" stands for the Kronecker product operation between matrices (vectors).
For a Hilbert space $\H$, its inner product is denoted by $\langle \cdot, \cdot \rangle_{\H}$; its norm induced by the inner product is defined by $\| \cdot \|_{\H} = \sqrt{\langle \cdot, \cdot \rangle_{\H}}$;
$\Phi^*: \H \rightarrow \R^p$ represents the adjoint operator of an operator $\Phi: \R^p \rightarrow \H$. 
$\E[\cdot]$ is the mathematical expectation. $\N(\bar \vx, \mP)$ denotes the Gaussian distribution with mean $\bar \vx$ and covariance matrix $\mP$.

\section{Preliminaries and Problem Formulation}\label{sec:problem formulation}

\subsection{Communication Topology}

The communication topology of the sensor network is modeled by an undirected graph $\G = (\V, \E, \mA)$, where $\V$, $\E$ and $\mA$ stand for the node set consisting of $\nS$ sensors, edge set consisting of bidirectional communication links between the sensors and weighted adjacency matrix, respectively.
For $i,j \in \V$, $\mA = [a_{ij}]$ is assumed to be double stochastic and satisfies $a_{ii} > 0$, $a_{ij} \geq 0$, $\sum_{j \in \V} a_{ij} = 1$, $\sum_{i \in \V} a_{ij} = 1$. The neighboring sensors of Node $i$ form a set $\N_i = \{j \in \V | a_{ij} > 0\}$ and $i \in \N_i$, from which Node $i$ receives information. 

\begin{assumption}\label{assum:topology}
	$\G$ is undirected and connected. 
\end{assumption}

Assumption  \ref{assum:topology} is standard and commonly used in \cite{Olfati-Saber2005Consensus,Olfati-Saber2005Distributed,Olfati-Saber2007Distributed}.  If Assumption \ref{assum:topology} satisfies, then for any pair of nodes ($i, j$), there is a path made up of edges between them.

\subsection{Distribution Embedding}

An RKHS, denoted by $\H$, on $\Omega$ is a Hilbert space of functions $g: \Omega \rightarrow \R$ equipped with a kernel function $k: \Omega \times \Omega \rightarrow \R$ satisfying
the following reproducing property:
\begin{align}
\langle \mg, \mk(\cdot, \vx) \rangle_{\H} &= \mg(\vx), \ \forall \mg \in \H,
\end{align}
and consequently,
\begin{align}
\langle \mk(\cdot, \vx), \mk(\cdot, \vx') \rangle_{\H} = \mk(\vx', \vx) = \mk(\vx, \vx').
\end{align}
The kernel function can also be treated as an implicit feature map $\phi: \Omega \rightarrow \H$ satisfying
\begin{align}
\langle \phi(\vx), \phi(\vx') \rangle_{\H} &= \mk(\vx', \vx).
\end{align}
Some commonly-used kernel functions on $\R^p$ are listed below for later use:
\begin{align}
k(\vx, \vx') &= (\langle x, x' \rangle_{\H} + c)^d, \ (\textrm{polynomial kernel}) \\
k(\vx, \vx') &= \exp \left(-\frac{\| \vx - \vx' \|^2}{ \sigma}\right), \ (\textrm{Gaussian kernel}) \label{eq:Gaussian kernel}\\
k(\vx, \vx') &=  \exp \left(-\frac{\| \vx - \vx' \|_1}{\sigma}\right),
(\textrm{Laplace kernel}) \label{eq:Laplace kernel}
\end{align}
where $c, \sigma > 0$,  $d \in \Na^+$.

The KME of a distribution with probability density function $p(\vx)$ is an element in the RKHS defined by
\begin{align}
\mu_{\vx} \coloneqq  \E_{\vx}[\phi(\vx)] 
= \int_{\Omega} \phi(\vx) p(\vx) \diff \vx,
\end{align}
and has the following property:
\begin{align}
	\langle \mu_{\vx}, \mg \rangle_{\H} \coloneqq \E_{\vx} [\mg(\vx)], \ \forall \mg \in \H.
\end{align}
Given a weighted sample $\{(\vx_l, w_l)\}_{l = 1}^{\nP}$ of $p(\vx)$,  the KME is approximated as
\begin{align}
	\sizehat \mu_{\vx} = \sum_{l = 1}^{\nP} w_l \phi(\vx_l).
\end{align}

For two separable Hilbert spaces $\F$ and $\G$ with $\{e_i\}_{i \in I}$ and $\{f_j\}_{j \in J}$ being their orthonormal bases, respectively, where index sets $I$ and $J$ are either finite or countably finite, the Hilbert--Schmidt norm of a compact linear operators $\mL: \G \rightarrow \F$ is defined as
\begin{align}\label{eq:Hilbert--Schmidt norm}
	\| \mL \|_{\HS} &= \sum_{j \in J} \| \mL f_j \|_{\F}^2\notag\\
	&= \sum_{i \in I} \sum_{j \in J} | \langle \mL f_j, e_i \rangle_{\F} |^2.
\end{align}
We say that the operator $\mL$ is Hilbert--Schmidt when \eqref{eq:Hilbert--Schmidt norm} is finite. For more details, one can refer to \cite{Song2013}. 

\subsection{Problem Formulation}

For the $i$-th node ($i = 1, 2, \dots, \nS$), the following nonlinear dynamic system is addressed:
\begin{equation}
\begin{aligned}\label{eq:state space model}
\vx_k &= \mf_{k - 1} (\vx_{k - 1}, \vw_k),\\
\vy_{i,k} &= \mh_{i,k} (\vx_k) + \vv_{i,k},
\end{aligned}
\end{equation}
where $\vx_k \in \R^{\dx}$ and $\vy_{i,k} \in \R^{\dy}$ are the state and the $i$-th node's measurement, respectively, $k = 1, 2, \dots$ is time index, $\mf_k: \R^{\dx} \rightarrow \R^{\dx}$ and $\mh_{i,k}: \R^{\dx} \rightarrow \R^{\dy}$ are known vector-valued transition and measurement functions for all $ k,i$, and typically nonlinear, the process noise $\{\vw_k\}$ and measurement noise $\{\vv_{i,k}\}$ are mutually independent zero-mean white noise sequences
whose joint covariance matrix is given by
\begin{align}
\E \bigg\{ \begin{bmatrix} \vw_k\\ \vv_{i,k}\end{bmatrix} [\vw_{k'}^{\Trans}\  \vv_{j,k'}^{\Trans}] \bigg\} = \begin{bmatrix}
\mQ_k  & 0\\
0      & \mR_{i,k} \delta_{i,j}
\end{bmatrix}
\delta_{k,k'},
\end{align}
where $\mQ_k \succeq 0$, $\mR_{i,k} \succ 0$, and $\delta_{i,j}$ is the Kronecker delta function.
The initial state with mean $\bar \vx_0$ and covariance matrix $\mP_0$ is independent of the noise.

Denote by $\vx_{k_1 | k_2}$ and $\mP_{k_1 | k_2}$ respectively an estimate of $\vx_{k_1}$ and its error covariance matrix, given the measurement sequences up to time $k_2$. The aim of filtering is to obtain $\vx_{k | k}$ and its error covariance matrix $\mP_{k | k}$ for the $k$. The optimal solution to such a filtering problem under the system \eqref{eq:state space model} is in the form of the centralized Bayesian filtering equations \cite{Mahler2014Advances}:
\begin{align}
p(\vx_k | \Y_{k - 1}) &= \int_{\R^{\dx}} p(\vx_k | \vx_{k - 1}) p(\vx_{k - 1} | \Y_{k - 1}) \diff \vx_{k - 1}, \\
p(\vx_k | \Y_k) &= c \cdot p(\vy_k | \vx_k) p(\vx_k | \Y_{k - 1}),
\end{align}
where $\Y_k = \{\vy_1, \dots, \vy_k\}$, $\vy_k = [\vy_{1,k}^{\Trans}, \dots, \vy_{\nS, k}^{\Trans}]^{\Trans}$ and $c$ is a normalization constant.
However, closed expressions for the prediction and posterior probability density functions, denoted by $p(\vx_k | \Y_{k - 1})$ and $p(\vx_k | \Y_k)$, respectively, are generally difficult or even impossible to obtain due to the high nonlinearity of the transition and measurement functions. Therefore, approximations are necessary in a practical sense.

In this paper, we aim at estimating the KME of the prediction and posterior probability density functions and then providing an effective distributed nonlinear filtering in the RKHS.


\section{Distributed Nonlinear Filter in RKHS}\label{sec:DNF}

The state is embedded into an RKHS $\H$ of functions with a feature map $\phi$ (time-invariant) defined by a positive-definite kernel function $k: \R^{\dx} \times \R^{\dx} \rightarrow \R$, i.e.,
\begin{align}
\phi(\vx) = \mk(\cdot, \vx).
\end{align}
For each node $i \in \V$, suppose a sample of the posterior distribution of state at time $k - 1$ is
\begin{align}\label{eq:node:post:sample}
\mX_{i, k - 1 | k - 1} = \left[\valpha_1, \dots, \valpha_{\nP}\right],
\end{align}
with corresponding weight vector
\begin{align}\label{eq:node:post:sample weight}
\vw = \left[w_1, \dots, w_{\nP}\right]^{\Trans},
\end{align}
where $\vones^{\Trans} \vw = 1$, and each $w_l \geq \epsilon$ for $l = 1, \dots, \nP$ with $\epsilon$ being a small positive real number. Then, we consider the prediction and update steps of filtering at time $k$ only, since it can be done recursively.

\subsection{Prediction}

In the prediction step, we aim at estimating the KME of the prediction distribution and its error covariance operator, denoted by $\mu_{i, k | k - 1}$ and $\mP_{i ,k | k - 1}^{\phi}$, respectively.
At time $k$, a sample of the prediction distribution is generated in accordance with the state transition equation in \eqref{eq:state space model}:
\begin{equation}\label{eq:predic:sample}
\begin{aligned}
\mX_{i, k | k - 1} &= \left[\vbeta_1, \dots, \vbeta_{\nP}\right], \\
\vbeta_l &\sim  \p(\vx_k | \valpha_l), \ l = 1, \dots, \nP,
\end{aligned}
\end{equation}
whose weight vector is given by \eqref{eq:node:post:sample weight}.
The predicted sample points $\{\vbeta_l\}_{l = 1}^{\nP}$ are generated by one of the following methods:
\begin{enumerate}[i)]
\item Deterministic sampling, such as the quasi Monte Carlo method (see, e.g., \cite{Lemieux2009}).
\item Stochastic sampling, such as the Monte Carlo method, where all nodes use the same random number generator with the same initial random number seed.
\end{enumerate}
Specifically, by generating sample points from the distribution of process noise $\vw_k$ and propagating them correspondingly through nonlinear functions $\mf_{k - 1}(\valpha_l, \cdot)$ ($l = 1, \dots, \nP$), we obtain $\{\vbeta_l\}_{l = 1}^{\nP}$. The commonality of the above two sampling methods lies in their fixed sample points for a given distribution. In another word, if all nodes have the same posterior sample $\mX_{i,k - 1|k - 1}$, then they have the same predicted sample $\mX_{i,k | k - 1}$ as well.

\begin{remark}
The prediction step changes the positions of the sample points of the posterior distribution while keeping the weights unchanged.
\end{remark}

An estimate of the KME of the prediction distribution is given by
\begin{align}
\mu_{i, k | k - 1} &= \sum_{l = 1}^{\nP} w_l \phi(\vbeta_l), \label{eq:RKHS:prediction:mean}
\end{align}
with its error covariance operator defined as
\begin{align}
\mP_{i ,k | k - 1}^{\phi} &= \sum_{l = 1}^{\nP} w_l \big(\phi(\vbeta_l) - \mu_{i, k | k - 1}\big)
\big(\phi(\vbeta_l) - \mu_{i, k | k - 1}\big)^{\Trans}.\label{eq:RKHS:prediction:cov}
\end{align}

\begin{remark}
If $k(\cdot, \cdot)$ is a bounded characteristic kernel, then $\mu_{i, k | k - 1}$ given by \eqref{eq:RKHS:prediction:mean} is a consistency estimate of the KME of the prediction distribution under weak assumptions \cite{Kanagawa2014Monte}.
$\mP_{i ,k | k - 1}^{\phi}$ is essentially a Hilbert--Schmidt linear operator mapping from $\H$ to itself. We use it to define the gain operator in Section \ref{sec:optimality} later for convergence analysis.
\end{remark}

Let
\begin{align}
\Phi &= \left[\phi(\vbeta_1), \dots, \phi(\vbeta_{\nP})\right],\label{eq:Phi}\\
\mW &= \diag(\vw) - \vw \vw^{\Trans}.\label{eq:W}
\end{align}
Then, we have the following proposition.

\begin{proposition}
The following equality holds:
\begin{align}
\mP_{i, k | k - 1}^{\phi} = \Phi \mW \Phi^*. \label{eq:error covariance operator:dis}
\end{align}
\end{proposition}

\begin{IEEEproof}
By some simple calculations, we have
\begin{align}
&\mP_{i, k | k - 1}^{\phi} \notag \\
&= (\Phi - \Phi \vw \vones^{\Trans}) \diag(\vw) (\Phi - \Phi \vw \vones^{\Trans})^{\Trans} \notag \\
&= \Phi (\mI - \vw \vones^{\Trans}) \diag(\vw) (\mI - \vones \vw^{\Trans}) \Phi^* \notag \\
&= \Phi \big( \diag(\vw) + \vw \vones^{\Trans} \diag(\vw) \vones \vw^{\Trans} - 2 \vw \vones^{\Trans} \diag(\vw) \big) \Phi^* \notag \\
&= \Phi \big( \diag(\vw) + \vw \vw^{\Trans} \vones \vw^{\Trans} - 2 \vw \vw^{\Trans} \big) \Phi^* \notag \\
&= \Phi \big( \diag(\vw) - \vw \vw^{\Trans} \big) \Phi^* \notag \\
&= \Phi \mW \Phi^*.
\end{align}
The proof is completed.
\end{IEEEproof}

\begin{remark}
	If $w_l \geq 0$ for $l = 1, \dots, \nP$, then $\mW$ is positive semi-definite. Specifically, for all $\vz = [z_1, \dots, z_{\nP}]^{\Trans} \in \R^{\nP}$, we have
	\begin{align}
		\vz^{\Trans} \mW \vz &= \vz^{\Trans} \diag(\vw) \vz - \vz^{\Trans} \vw \vw^{\Trans} \vz \notag \\
		&= \sum_{l = 1}^{\nP} w_l z_l^2 - \Bigg(\sum_{l = 1}^{\nP} w_l z_l \Bigg)^2 \notag\\
		&\geq 0,
	\end{align}
	where the inequality holds by Jensen's inequality (see Page 246 of \cite{Zorich2015Mathematical}).
\end{remark}

\subsection{Update}

In the update step, we aim at obtaining the KME and a sample of the posterior distribution, denoted by $\mu_{i, k | k}$ and $\mX_{i, k | k}$, respectively.
The main difficulty of update lies in dealing with the nonlinearity of the measurement functions.
For $i \in \V$, we care about the nonlinearity of $\mh_{i,k}(\cdot)$ in the range of the state.
By representing a real-valued function as an inner product in RKHS \cite{Muandet2017}, we express $\mh_{i,k}(\cdot)$ in the following linear form:
\begin{align}
\mh_{i,k}(\cdot) &= \Big[\mh_{i,k}^{(1)}(\cdot), \dots, \mh_{i,k}^{(\dy)}(\cdot)\Big]^{\Trans} \notag\\
&\triangleq \Big[\big\langle \lambda^{(1)}_{i,k}, \phi(\cdot) \big\rangle_{\H}, \dots, \big\langle \lambda^{(\dy)}_{i,k}, \phi(\cdot) \big\rangle_{\H}\Big]^{\Trans} \notag\\
&\triangleq \vLambda_{i,k}^* \phi(\cdot),
\end{align}
where 
\begin{align}
\vLambda_{i,k} = \Big[\lambda^{(1)}_{i,k}, \dots, \lambda^{(\dy)}_{i,k}\Big] \label{eq:Lambda ik}
\end{align}
is a Hilbert--Schmidt linear operator mapping from $\R^{\dy}$ to $\H$, with $\lambda^{(1)}_{i,k}, \dots, \lambda^{(\dy)}_{i,k} \in \H$.
Then, the measurement equation in \eqref{eq:state space model} is converted into
\begin{equation}\label{system:linear:embedded}
\begin{aligned}
\vy_{i,k} &= \vLambda_{i,k}^* \phi(\vx_k) + \vv_{i,k}.
\end{aligned}
\end{equation}

It is worth emphasizing that $\vy_{i,k}$ is linear with respect to $\phi(\vx_k)$ since $ \vLambda_{i,k}$ is linear, but still nonlinear with respect to $\vx_k$ since $\phi$ is nonlinear. By expressing the nonlinear measurement function as inner products of a higher dimensional space (i.e., $\H$), the corresponding filtering problem can be simplified significantly while the original nonlinearity can be maintained.

\begin{remark}
By introducing a class of particularly large RKHSs, a nonlinear function can be approximated up to arbitrary accuracy in some sense under certain conditions (see Section 4.6 of \cite{Steinwart2008Support}). In fact, if $k(\cdot,\cdot)$ is a Gaussian kernel, then for all $\varepsilon > 0$, $q \in [1,\infty)$, there exists $\sizetilde \vLambda_{i,k}$ such that $\sizetilde \mh_{i,k}(\cdot) = \sizetilde \vLambda_{i,k}^* \phi(\cdot)$ satisfying
\begin{align}\label{eq:large Gaussian kernel}
\bigg(\int_{\R^{\dx}} \big\| \mh_{i,k}(\vx_k) - \sizetilde \mh_{i,k}(\vx_k) \big\|^q_q \cdot \p(\vx_k | \Y_{k - 1}) \diff \vx_k \bigg)^{\frac{1}{q}} &< \varepsilon \notag\\
\Leftrightarrow \int_{\R^{\dx}} \big\| \mh_{i,k}(\vx_k) - \sizetilde \mh_{i,k}(\vx_k) \big\|^q_q \cdot  \p(\vx_k | \Y_{k - 1}) \diff \vx_k &< \delta,
\end{align}
where $\delta \geq \varepsilon^q > 0$.
\end{remark}

A sample of noiseless measurement for Node $i$ at time $k$ is
\begin{equation}\label{eq:sample:noiseless measurement:node}
\begin{aligned}
\mY_{i,k} = \big[\mh_{i,k}(\vbeta_1), \dots, \mh_{i,k}(\vbeta_{\nP})\big],
\end{aligned}
\end{equation}
with weight vector given by \eqref{eq:node:post:sample weight}.
Then, an estimate of $\vLambda_{i,k}$ is determined by solving the following minimization problem:
\begin{align}\label{min:node:SLR}
\sizehat \vLambda_{i,k} = \underset{\vLambda}{\arg \min} \ \sum_{l = 1}^{\nP} w_l \big\| \mh_{i,k}(\vbeta_l) - \vLambda^* \phi(\vbeta_l) \big\|^2.
\end{align}
Note that the objective function of \eqref{min:node:SLR} is essentially an estimate of the second integral in \eqref{eq:large Gaussian kernel} by taking $q = 2$. This is a convex optimization problem whose optimal solution is as follows.

\begin{theorem}\label{thm:node:SLR}
The solution to \eqref{min:node:SLR} is
\begin{align}\label{eq:Lambda:i}
\sizehat \vLambda_{i,k} = \Phi \gram^{-1} \mY_{i,k}^{\Trans},
\end{align}
where $K$ is the Gram matrix (see Page 117 of \cite{Steinwart2008Support}) given by
\begin{align}\label{eq:gram matrix}
\gram =
\begin{bmatrix}
\mk(\vbeta_1, \vbeta_1) & \dots & \mk(\vbeta_1, \vbeta_{\nP}) \\
\vdots          & \ddots & \vdots                  \\
\mk(\vbeta_{\nP}, \vbeta_1) & \dots & \mk(\vbeta_{\nP}, \vbeta_{\nP})
\end{bmatrix}.
\end{align}
\end{theorem}

\begin{IEEEproof}
Let $\vLambda = [\lambda_1, \dots, \lambda_{\dy}]$, where $\lambda_s \in \H$, with $s = 1, \dots, \dy$. Then, benefited from the representer theorem \cite{Schoelkopf2001}, each $\lambda_s$ can be expressed as
\begin{align}
\lambda_s = \sum_{l = 1}^{\nP} b_l \phi(\vbeta_l), \ s = 1, \dots, \dy,
\end{align}
where $b_1, \dots, b_{\nP}$ are the coefficients.
Then, $\vLambda$ can be written as
\begin{align}
\vLambda &=
\begin{bmatrix}
\sum_{l = 1}^{\nP} b_{l,1} \phi(\vbeta_l) & \dots & \sum_{l = 1}^{\nP} b_{l,\dy} \phi(\vbeta_l)
\end{bmatrix}
= \Phi \mB,
\end{align}
with
\begin{align}
\mB = \begin{bmatrix}
b_{11} & \cdots & b_{1 \dy} \\
\vdots      & \ddots & \vdots       \\
b_{\nP 1} & \dots & b_{\nP \dy}
\end{bmatrix}.
\end{align}
Denoting $\sqrt \vw = [\sqrt{w_1}, \dots, \sqrt{w_{\nP}}]^{\Trans}$, and $\mV = \diag(\sqrt \vw)$,
we rewrite \eqref{min:node:SLR} in a more comprehensive fashion as
\begin{align}\label{min:compre}
\underset{\mB}{\min} \ \| \mY_{i,k} \mV - \mB^{\Trans} \gram \mV \|^2.
\end{align}
Then, the solution to \eqref{min:node:SLR} is given by $\sizehat \vLambda_{i,k} = \Phi \mB$.
The derivative of the objective function of \eqref{min:compre} with respect to $\mB$ is
\begin{align}
&-2 \gram \mV \mV \mY_{i,k}^{\Trans} + 2 \gram \mV \mV \gram \mB.
\end{align}
Setting the derivative to zero, we have
\begin{align}
\gram \mV \mV \gram \mB = \gram \mV \mV \mY_{i,k}^{\Trans}.
\end{align}
Then, the solution to \eqref{min:compre} is
\begin{align}
\mB &= (\gram \mV \mV \gram)^{-1} \gram \mV \mV \mY_{i,k}^{\Trans} \notag\\
&= \big(\gram \diag(\vw) \gram\big)^{-1} \gram \diag(\vw) \mY_{i,k}^{\Trans} \notag\\
&= \gram^{-1} \mY_{i,k}^{\Trans},
\end{align}
and thus, $\sizehat \vLambda_{i,k} = \Phi \gram^{-1} \mY_{i,k}^{\Trans}$.
\end{IEEEproof}

From Theorem \ref{thm:node:SLR}, we know that the minimizer of \eqref{min:node:SLR} is independent of the weight vector $\vw$. This is because $\mh_{i,k}(\cdot)$ is fitted exactly by $\sizehat \vLambda_{i,k}^* \phi(\cdot)$ in the evaluated points $\vbeta_1, \dots, \vbeta_{\nP}$.

\begin{remark}
If the kernel function $k(\cdot, \cdot)$ is strictly positive definite, then the Gram matrix $\gram$ is positive definite. In practice, $\gram^{-1}$  can be replaced by $(\gram + \sigma \mI)^{-1}$ with $\sigma > 0$ being a small real number.
\end{remark}

We then provide the following update rule to obtain the KME of the posterior distribution of state:
\begin{align}\label{eq:KME:node:post}
\mu_{i, k | k} = \mu_{i, k | k - 1} + \Phi \mW^{\frac{1}{2}} \vGamma_{i,k}^{-1} \vxi_{i,k},
\end{align}
where the matrix $\vGamma_{i,k} \in \R^{\nP \times \nP}$ and the vector $\vxi_{i,k} \in \R^{\nP}$ are determined by Algorithm \ref{alg:ACF}. For $i \in \V$, the initialization is designed as follows:
\begin{align}
\vGamma_{i,k}^{(0)} &= \nS \mW^{\frac{1}{2}} \mY_{i,k}^{\Trans} \mR_{i,k}^{-1} \mY_{i,k} \mW^{\frac{1}{2}} + \mI, \\
\vxi_{i,k}^{(0)} &= \nS \mW^{\frac{1}{2}} \mY_{i,k}^{\Trans} \mR_{i,k}^{-1} \big(\vy_{i,k} - \sizehat \vLambda_{i,k}^* \mu_{i,k | k - 1}\big).
\end{align}

\begin{algorithm}[H]
\renewcommand{\algorithmicrequire}{\textbf{Input:}}
\renewcommand{\algorithmicensure}{\textbf{Output:}}
\caption{Distributed Calculation for $\vGamma_{i,k}$ and $\vxi_{i,k}$}
\label{alg:ACF}
\begin{algorithmic}[1]
\REQUIRE $\vGamma_{j,k}^{(0)}$ and $\vxi_{j,k}^{(0)}$, for all $j \in \N_i$
\STATE Set $r = 1$.
\REPEAT
\STATE Implement
\begin{align}
\vGamma_{i,k}^{(r)} &= \vGamma_{i,k}^{(r - 1)} + \sum_{j \in \N_i} a_{ij} \left(\vGamma_{j,k}^{(r - 1)} - \vGamma_{i,k}^{(r - 1)}\right), \label{eq:iteration:ACF:gamma} \\
\vxi_{i,k}^{(r)} &= \vxi_{i,k}^{(r - 1)} + \sum_{j \in \N_i} a_{ij} \left(\vxi_{j,k}^{(r - 1)} - \vxi_{i,k}^{(r - 1)}\right).\label{eq:iteration:ACF:xi}
\end{align}
\STATE Set $r = r + 1$.
\UNTIL convergence to get $\vGamma_{i,k}^*$, $\vxi_{i,k}^*$.
\ENSURE  $\vGamma_{i,k}^*$, $\vxi_{i,k}^*$
\end{algorithmic}
\end{algorithm}

Set
\begin{align}
	\vGamma_k^{(r)} &= \begin{bmatrix}
		\vGamma_{1,k}^{(r) \Trans}, & \vGamma_{2,k}^{(r) \Trans}, & \dots, & \vGamma_{\nS,k}^{(r) \Trans}
	\end{bmatrix}^{\Trans}, \\
	\vxi_k^{(r)} &= \begin{bmatrix}
		\vxi_{1,k}^{(r) \Trans}, & \vxi_{2,k}^{(r) \Trans}, & \dots, & \vxi_{\nS,k}^{(r) \Trans}
	\end{bmatrix}^{\Trans}.
\end{align}
Then, \eqref{eq:iteration:ACF:gamma} and \eqref{eq:iteration:ACF:xi} can be written more compactly as
\begin{align}
\vGamma_k^{(r)} &= (\mA \otimes \mI) \vGamma_k^{(r - 1)}
= (\mA^r \otimes \mI) \vGamma_k^{(0)}, \\
\vxi_k^{(r)} &= (\mA \otimes \mI) \vxi_k^{(r - 1)}
= (\mA^r \otimes \mI) \vxi_k^{(0)}.
\end{align}
From \cite{Xiao2004Fast}, we know that if Assumption \ref{assum:topology} holds, then for all $i \in \V$, the following two equations hold:
\begin{align}
\lim_{r \rightarrow \infty} \vGamma_{i,k}^{(r)} &= \frac{1}{\nS} \sum_{j \in \V} \vGamma_{j,k}^{(0)} \notag\\
&= \frac{1}{\nS} \sum_{j \in \V} \big(\nS \mW^{\frac{1}{2}} \mY_{j,k}^{\Trans} \mR_{j,k}^{-1} \mY_{j,k} \mW^{\frac{1}{2}} + \mI\big), \label{eq:ACF:converge:gamma}\\
\lim_{r \rightarrow \infty} \vxi_{i,k}^{(r)} &= \frac{1}{\nS} \sum_{j \in \V} \vxi_{j,k}^{(0)} \notag\\
&= \frac{1}{\nS} \sum_{j \in \V} \nS \mW^{\frac{1}{2}} \mY_{j,k}^{\Trans} \mR_{i,k}^{-1} \big(\vy_{j,k} - \sizehat \vLambda_{j,k}^* \mu_{j,k | k - 1}\big).\label{eq:ACF:converge:xi}
\end{align}
Clearly, the right hand sides of \eqref{eq:ACF:converge:gamma} and \eqref{eq:ACF:converge:xi} are free of index $i$. This means that each node achieves the average consensus by communicating with each other.

From \eqref{eq:RKHS:prediction:mean}, we know that $\mu_{i, k | k - 1} = \Phi \vw$. Then, \eqref{eq:KME:node:post} can be rewritten as
\begin{align}\label{eq:KME:update:weight}
\mu_{i, k | k} = \Phi (\vw + \mW^{\frac{1}{2}} \vGamma_{i,k}^{-1} \vxi_{i,k}).
\end{align}
Next, we prove that $\mu_{i, k | k}$ is a weighted combination of each column of $\Phi$.
Letting
\begin{align}\label{eq:nu}
\vnu = \vw + \mW^{\frac{1}{2}} \vGamma_{i,k}^{-1} \vxi_{i,k},
\end{align}
we have the following theorem.

\begin{theorem}\label{thm:post weight}
$\vnu$ is a weight vector satisfying $\vones^{\Trans} \vnu = 1$.
\end{theorem}

\begin{IEEEproof}
Let
$\sizetilde \mY_{i,k} = \mY_{i,k} \mW^{\frac{1}{2}}$.
By some calculations, we have
\begin{align}
\vones^{\Trans} \vnu
&= \vones^{\Trans} (\vw + \mW^{\frac{1}{2}} \vGamma_{i,k}^{-1} \vxi_{i,k}) \notag\\
&= 1 + \vones^{\Trans} \mW^{\frac{1}{2}} \vGamma_{i,k}^{-1} \vxi_{i,k} \notag\\
&= 1 + \vones^{\Trans} \mW^{\frac{1}{2}} \Bigg(\frac{1}{\nS} \sum_{i \in \V} \big(\nS \sizetilde \mY_{i,k}^{\Trans} \mR_{i,k}^{-1} \sizetilde \mY_{i,k} + \mI \big)\Bigg)^{-1} \notag\\
&\quad \cdot \frac{1}{\nS} \sum_{i \in \V} \nS \sizetilde \mY_{i,k}^{\Trans} \mR_{i,k}^{-1} (\vy_{i,k} - \sizehat \vLambda_{i,k}^* \mu_{i,k | k - 1}) \notag\\
&= 1 + \vones^{\Trans} \mW^{\frac{1}{2}} \Bigg(\sum_{i \in \V} \sizetilde \mY_{i,k}^{\Trans} \mR_{i,k}^{-1} \sizetilde \mY_{i,k} + \mI \Bigg)^{-1} \notag\\
&\quad \cdot \sum_{i \in \V} \sizetilde \mY_{i,k}^{\Trans} \mR_{i,k}^{-1} (\vy_{i,k} - \sizehat \vLambda_{i,k}^* \mu_{i,k | k - 1}). \label{eq:weight:zero}
\end{align}
Denote
\begin{align}
\sizetilde \mY_k &= \big[\sizetilde \mY_{1,k}^{\Trans}, \dots, \sizetilde \mY_{\nS,k}^{\Trans}\big]^{\Trans}, \\
\mR_k &= \blkdiag(\mR_{1,k}, \dots, \mR_{\nS,k}).
\end{align}
Then, using Woodbury matrix identity (see Page 258 of \cite{Higham2002Accuracy}), the right hand side of \eqref{eq:weight:zero} is equivalent to
\begin{align}
& 1 + \vones^{\Trans} \mW^{\frac{1}{2}} \Bigg(\mI - \sizetilde \mY_k^{\Trans} \Big(\mR_k + \sizetilde \mY_k \sizetilde \mY_k^{\Trans}\Big)^{-1} \sizetilde \mY_k \Bigg) \notag \\
&\quad \cdot \sum_{i \in \V} \sizetilde \mY_{i,k}^{\Trans} \mR_{i,k}^{-1} (\vy_{i,k} - \sizehat \vLambda_{i,k}^* \mu_{i,k | k - 1}) \notag\\
&= 1 + \Bigg(\vones^{\Trans} \mW^{\frac{1}{2}} - \vones^{\Trans} \mW^{\frac{1}{2}} \sizetilde \mY_k^{\Trans} \Big(\mR_k + \sizetilde \mY_k \sizetilde \mY_k^{\Trans}\Big)^{-1} \sizetilde \mY_k \Bigg) \notag\\
&\quad \cdot \mW^{\frac{1}{2}} \sum_{i \in \V} \mY_{i,k}^{\Trans} \mR_{i,k}^{-1} (\vy_{i,k} - \sizehat \vLambda_{i,k}^* \mu_{i,k | k - 1}) \notag\\
&= 1 + \vones^{\Trans} \mW \sum_{i \in \V} \mY_{i,k}^{\Trans} \mR_{i,k}^{-1} (\vy_{i,k} - \sizehat \vLambda_{i,k}^* \mu_{i,k | k - 1}) \notag\\
&= 1.
\end{align}
Here, we use the fact that
\begin{align}
\vones^{\Trans} \mW &= \vones^{\Trans} \diag(\vw) - \vones^{\Trans} \vw \vw^{\Trans} \notag\\
&= \vw^{\Trans} - \vw^{\Trans} \notag\\
&= 0.
\end{align}
This completes the proof.
\end{IEEEproof}

Based on \eqref{eq:KME:update:weight}, \eqref{eq:nu} and Theorem \ref{thm:post weight}, we immediately obtain a sample of the posterior distribution
\begin{equation}
\begin{aligned}\label{eq:weight:posterior sample}
\mX_{i, k | k} &= \left[\vbeta_1, \dots, \vbeta_{\nP}\right] 
\end{aligned}
\end{equation}
with weight vector given by \eqref{eq:nu}.
The elements of $\vnu$ may be negative due to the randomness of $\vy_{i,k}$ ($i \in \V$). Thus, we normalize them by minimizing the empirical maximum mean discrepancy:
\begin{equation}\label{min:normalize:weight}
\begin{aligned}
&\underset{\sizetilde \vnu}{\min} \ \big\| \Phi \vnu - \Phi \sizetilde \vnu \big\|^2_{\H} \\
&\ \mathrm{s.t.} \  \vones^{\Trans} \sizetilde \vnu = 1,\\
&\qquad \ \; \,\sizetilde \nu_l \geq \epsilon, \ l = 1, \dots, \nP,
\end{aligned}
\end{equation}
where $\sizetilde \vnu = [\sizetilde \nu_1, \dots, \sizetilde \nu_{\nP}]^{\Trans}$, $\epsilon$ is a small positive real number. Note that
\begin{align}
\big\| \Phi \vnu - \Phi \sizetilde \vnu \big\|^2_{\H}
&= (\sizetilde \vnu - \vnu)^{\Trans} \Phi^* \Phi (\sizetilde \vnu - \vnu) \notag\\
&= (\sizetilde \vnu - \vnu)^{\Trans} \gram (\sizetilde \vnu - \vnu).
\end{align}
Then, the problem \eqref{min:normalize:weight} is a convex quadratic programming and can be solved efficiently by many classical numerical methods such as the interior-point methods \cite{Vanderbei1993Symmetric}.

\begin{remark}
In contrast to the $L_2$ distance between two kernel density estimates,
the maximum mean discrepancy between two embeddings is demonstrated to have more power against local departures from the null hypothesis for a two-sample test
\cite{sriperumbudur2010hilbert,Song2013}.
\end{remark}

\subsection{Filtering Algorithm}

For time $k = 1, 2, \dots$, each node $i$ implements the distributed nonlinear filter in RKHS (DNF-RKHS) summarized in Algorithm \ref{alg:filter}.

\begin{algorithm}[htbp]
\renewcommand{\algorithmicrequire}{}
\renewcommand{\algorithmicensure}{}
\caption{DNF-RKHS}
\label{alg:filter}
\begin{algorithmic}[1]
\REQUIRE \textbf{Input:} posterior sample $\mX_{i,k - 1,k - 1}$ at time $k - 1$, corresponding weight vector $\vw$
\REQUIRE \textbf{I. One-Step Prediction:}
\STATE Generate a sample of the prediction distribution by \eqref{eq:predic:sample}.
\STATE Calculate the KME of the prediction distribution by \eqref{eq:RKHS:prediction:mean}.
\REQUIRE \textbf{II. Update:}
\STATE Implement Algorithm \ref{alg:ACF} to get $\vGamma_{i,k}^*$, $\vxi_{i,k}^*$.
\STATE Calculate the KME of the posterior distribution by \eqref{eq:KME:node:post}.
\STATE Take a sample of the posterior distribution by \eqref{eq:weight:posterior sample}.

\STATE Update the weights of the posterior sample by solving \eqref{min:normalize:weight}.
\ENSURE \textbf{Output:} $\mu_{i, k | k}$, posterior sample $\mX_{i,k | k}$ at time $k$, corresponding weight vector $\sizetilde \vnu$
\end{algorithmic}
\end{algorithm}

\begin{remark}
The update step changes the weights of the sample points of the prediction distribution while keeping the positions unchanged. This is similar to the celebrated particle filter where the positions of the particles are changed to fit the prediction distribution in the prediction step and the weights are updated to fit the posterior distribution in the update step.
\end{remark}

We consider the pre-image problem of the DNF-RKHS, that is, to recover the state estimation in the state space. At time $k$, a sample of the posterior distribution can be obtained after executing Algorithm \ref{alg:filter}. Thus, we can obtain the desired moment approximations. In the minimum mean squared error sense, the optimal state estimate and its error covariance matrix are given by the mean and covariance matrix of the posterior distribution, respectively, which are calculated as follows:
%
\begin{align}
\vx_{i, k |k} &= \sum_{l = 1}^{\nP} \sizetilde \nu_l \vbeta_l,  \\
\mP_{i, k | k} &= \sum_{l = 1}^{\nP} \sizetilde \nu_l (\vbeta_l - \vx_{i, k |k}) (\vbeta_l - \vx_{i, k |k})^{\Trans}.
\end{align}

\begin{remark}
Generally, the pre-image problem of recovering a state distribution in state space model from its KME is not easy, as it is difficult to get a sample of the distribution from its KME directly.
Benefitted from the posterior sample points we get in Algorithm \ref{alg:filter}, the corresponding pre-image problem becomes simple.
\end{remark}


\subsection{Implementation Issues}

In implementing Algorithm \ref{alg:filter}, the weights of the sample of the posterior distribution may be assigned to a few points and thus make many points useless.
Technically, an effective way of avoiding this phenomenon is resampling. By repeatedly drawing randomly from a discrete approximation of the posterior probability density function:
\begin{align}
	p(\vx_k | \Y_k) \approx \sum_{l = 1}^{\nP} \sizetilde \nu_l \delta (\vx_k - \vbeta_l),
\end{align}
where $\delta(\cdot)$ is the Dirac delta function, the new sample can be created.
It is independent and identically distributed and uniformly weighted. Although the weights still accumulate on the points with large probabilities, in the subsequent prediction step, the same posterior sample point will correspond to different predicted sample points resulting from the process noise.

It is important to save communication cost in a distributed filtering algorithm.  
From Algorithm~\ref{alg:ACF}, we know that at each time $k$, for $i \in \V$, $j \in \N_i$, an $\nP \times \nP$ matrix $\Gamma_{j,k}$ and an $\nP$ dimensional vector $\vxi_{j,k}$ should be transmitted from the $j$-th sensor to the $i$-th sensor. If the sample size $\nP$ is large, then so is the communication cost. In contrast, transmitting an $\dy \times \nP$ matrix $\mY_{j,k}$, an $\dy \times \dy$ matrix $\mR_{j,k}$ and an $\dy$ dimensional vector $\vy_{j,k}$ is more efficient.

\section{Convergence Analysis}\label{sec:optimality}

In this section, we prove that the proposed DNF-RKHS converges to a centralized filtering algorithm as $r \rightarrow \infty$. 

First, we provide a centralized filtering algorithm, namely, centralized nonlinear filter in RKHS (CNF-RKHS), consisting of recursive prediction and update of the KMEs of the state distributions.
In the central location, the sample points of the initial state distribution are generated and the prediction step are done in the same way as those in the DNF-RKHS, i.e.,
\begin{align}
\mX_{0 | 0} &= \mX_{i, 0 | 0}, \\
\mu_{k | k - 1} &= \mu_{i, k | k - 1}, \label{eq:KME:pre den:dis and cen}\\
\mP_{k | k - 1}^{\phi} &= \mP_{i ,k | k - 1}^{\phi}. \label{eq:error cov oper:cen}
\end{align}
Once receiving the measurements from all sensors, set
\begin{align}
\vy_k = \begin{bmatrix}\vy_{1,k} \\ \vdots \\ \vy_{\nS,k}\end{bmatrix},\
\mh_k(\cdot) = \begin{bmatrix} \mh_{1,k}(\cdot) \\ \vdots \\ \mh_{\nS,k}(\cdot) \end{bmatrix},\
\vv_k = \begin{bmatrix}\vv_{1,k} \\ \vdots \\ \vv_{\nS,k}\end{bmatrix},
\end{align}
where $\vy_k \in \R^{\dY}$ is augmented measurement vector, $\dY = \sum_{i = 1}^{\nS} \dy$, $\mh_k: \R^{\dx} \rightarrow \R^{\dY}$ is augmented measurement function,
augmented noise $\vv_k \in \R^{\dY}$ has covariance matrix
\begin{align}
\mR_k = \blkdiag(\mR_{1,k}, \dots, \mR_{\nS,k}).
\end{align}
Then, the centralized measurement equation is given by
\begin{equation}\label{system:measurement:centralized}
\begin{aligned}
\vy_k = \mh_k(\vx_k) + \vv_k.
\end{aligned}
\end{equation}
By embedding the state into $\H$, \eqref{system:measurement:centralized} is converted into
\begin{equation}\label{system:centralized:kernel:measurement}
\begin{aligned}
\vy_k = \vLambda_k^* \phi(\vx_k) + \vv_k,
\end{aligned}
\end{equation}
where
$\vLambda_k = \big[\lambda^{(1)}_k, \dots, \lambda^{(\dY)}_k \big]$ is a Hilbert--Schmidt linear operator from $\R^{\dY}$ to $\H$.
Similarly to \eqref{min:node:SLR}, an estimate of $\vLambda_k$ is determined by the following minimization problem:
\begin{align}\label{min:cen:SLR}
\sizehat \vLambda_k = \underset{\vLambda}{\arg \min} \ \sum_{l = 1}^{\nP} w_l \big\| \mh_k(\vbeta_l) - \vLambda^* \phi(\vbeta_l) \big\|^2.
\end{align}
Denote a sample of noiseless measurement for central location at time $k$ by
\begin{align}
\mY_k = \big[\mh_k(\vbeta_1), \dots, \mh_k(\vbeta_{\nP})\big],
\end{align}
with weight vector given by \eqref{eq:node:post:sample weight}. Then, the following result similar to Theorem \ref{thm:node:SLR} is given.

\begin{proposition}\label{prop:Lambda}
The solution to \eqref{min:cen:SLR} is
\begin{align}
\sizehat \vLambda_k = \Phi \gram^{-1} \mY_k^{\Trans}.\label{eq:lambda k}
\end{align}
\end{proposition}

\begin{IEEEproof}
The proof is similar to that of Theorem \ref{thm:node:SLR}, and so, omitted here.
\end{IEEEproof}

Then, we design the update rule of the KME as follows:
\begin{align}
\mu_{k | k} &= \mu_{k | k - 1} + \mG_k (\vy_k - \sizehat \vLambda_k^* \mu_{k | k - 1}), \label{eq:Kalman:update:mean}
\end{align}
with gain operator $\mG_k: \R^{\dY} \rightarrow \H$ given by
\begin{align}
\mG_k &= \mP^{\phi}_{k | k - 1} \sizehat \vLambda_k (\sizehat \vLambda^*_k \mP^{\phi}_{k | k - 1} \sizehat \vLambda_k +  \mR_k)^{-1}. \label{eq:Kalman:update:gain}
\end{align}
It is not difficult to see that such an update rule is essentially a direct extension of the standard Kalman filtering in the Euclidean space to the RKHS.

Generally, a ``good" nonlinear filter should degenerate to the optimal filter under a linear dynamic system. Specifically,
for linear dynamic system with additive noise:
\begin{equation}
\begin{aligned}\label{eq:linear state space model}
\vx_k &= \mF_{k - 1} \vx_{k - 1} + \vw_k,\\
\vy_{i,k} &= \mH_{i,k} \vx_k + \vv_{i,k},
\end{aligned}
\end{equation}
where $\mF_{k - 1} \in \R^{\dx \times \dx}$ and $\mH_{i,k} \in \R^{\dy \times \dx}$ are known matrices, we have the following results.

\begin{proposition}\label{prop:KF}
The CNF-RKHS degenerates to the centralized Kalman filter.
\end{proposition}

\begin{IEEEproof}
By taking $\phi$ as the identity map, \eqref{eq:Kalman:update:mean} with \eqref{eq:Kalman:update:gain} reduces to the standard Kalman filter.
\end{IEEEproof}


From Proposition \ref{prop:KF}, we know that the proposed CNF-RKHS extends the centralized Kalman filter to nonlinear dynamic systems by constructing recursive KMEs of state distributions in the higher-dimensional RKHS.

Next, we prove that the DNF-RKHS converges to the CNF-RKHS. Before that, the following two lemmas are necessary.

\begin{lemma}\label{lem:lambda}
	The following equality holds:
	\begin{align}
		\sizehat \vLambda_k = \big[\sizehat \vLambda_{1,k}, \dots, \sizehat \vLambda_{\nS,k} \big],\label{eq:Lambda:dis and cen}
	\end{align}
where $\sizehat \vLambda_k$ is given by \eqref{eq:lambda k}, and $\sizehat \vLambda_{i,k}$, $i \in \V$, are given by \eqref{eq:Lambda ik}.
\end{lemma}

\begin{IEEEproof}
	Note that
	\begin{align}
		\mY_k = \big[\mY_{1,k}^{\Trans}, \dots, \mY_{\nS,k}^{\Trans}\big]^{\Trans},
	\end{align}
	where each $\mY_{i,k}$ for $i \in \V$ corresponds to a sample of the noiseless measurement obtained at Node $i$, as given by \eqref{eq:sample:noiseless measurement:node}. Then, we have
	\begin{align}
		\sizehat \vLambda_k 
		&= \Phi \gram^{-1} \big[\mY_{1,k}^{\Trans}, \dots, \mY_{\nS,k}^{\Trans}\big] \notag\\
		&= \big[ \Phi \gram^{-1} \mY_{1,k}^{\Trans}, \dots, \Phi \gram^{-1} \mY_{\nS,k}^{\Trans} \big] \notag\\
		&= \big[\sizehat \vLambda_{1,k}, \dots, \sizehat \vLambda_{\nS,k} \big].
	\end{align}
	This completes the proof.
\end{IEEEproof}

\begin{lemma}\label{lemma:lambda and mu}
	The following equality holds:
	\begin{align}
		\sizehat \vLambda_k^* \mu_{k | k - 1} = \Big[\big(\sizehat \vLambda_{1,k}^* \mu_{1,k | k - 1}\big)^{\Trans}, \dots, \big(\sizehat \vLambda_{\nS,k}^* \mu_{\nS,k | k - 1}\big)^{\Trans}\Big]^{\Trans},
	\end{align}
where $\mu_{k | k - 1}$ is given by \eqref{eq:KME:pre den:dis and cen}, and $\mu_{i,k | k - 1}$, $i \in \V$, are given by \eqref{eq:RKHS:prediction:mean}.
\end{lemma}

\begin{IEEEproof}
	From \eqref{eq:KME:pre den:dis and cen} and \eqref{eq:Lambda:dis and cen}, we have
	\begin{align}
		\sizehat \vLambda_k^* \mu_{k | k - 1} &= \Big(\big[\sizehat \vLambda_{1,k}, \dots, \sizehat \vLambda_{\nS,k} \big]\Big)^* \mu_{k | k - 1} \notag \\
		&= \Big[\big(\sizehat \vLambda_{1,k}^* \mu_{k | k - 1}\big)^{\Trans}, \dots, \big(\sizehat \vLambda_{\nS,k}^* \mu_{k | k - 1}\big)^{\Trans}\Big]^{\Trans} \notag \\
		&= \Big[\big(\sizehat \vLambda_{1,k}^* \mu_{1,k | k - 1}\big)^{\Trans}, \dots, \big(\sizehat \vLambda_{\nS,k}^* \mu_{\nS,k | k - 1}\big)^{\Trans}\Big]^{\Trans}.
	\end{align}
	This completes the proof.
\end{IEEEproof}

Then, we have the following theorem.

\begin{theorem}\label{thm:converge}
The DNF-RKHS converges to the CNF-RKHS as $r \rightarrow \infty$.
\end{theorem}

\begin{IEEEproof}
From \eqref{eq:KME:node:post}, \eqref{eq:ACF:converge:gamma}, \eqref{eq:ACF:converge:xi}, and the fact that each node results in the same KME of the prediction distribution, as $r \rightarrow \infty$, the KME of the posterior distribution for each node converges to
\begin{align}
\mu_{i, k | k} = \mu_{k | k - 1} + \Phi \mW^{\frac{1}{2}} \vGamma_k^{-1} \vxi_k,
\end{align}
where
\begin{align}
\vGamma_k &= \frac{1}{\nS} \sum_{i \in \V} \big(\nS \mW^{\frac{1}{2}} \mY_{i,k}^{\Trans} \mR_{i,k}^{-1} \mY_{i,k} \mW^{\frac{1}{2}} + \mI\big), \\
\vxi_k &= \frac{1}{\nS} \sum_{i \in \V} \nS \mW^{\frac{1}{2}} \mY_{i,k}^{\Trans} \mR_{i,k}^{-1} \big(\vy_{i,k} - \sizehat \vLambda_{i,k}^* \mu_{i,k | k - 1}\big).
\end{align}
By some calculations, we have
\begin{align}
&\Phi \mW^{\frac{1}{2}} \vGamma_k \vxi_k \notag\\
&= \Phi \mW^{\frac{1}{2}} \Bigg( \sum_{i \in \V} \mW^{\frac{1}{2}} \mY_{i,k}^{\Trans} \mR_{i,k}^{-1} \mY_{i,k} \mW^{\frac{1}{2}} + \mI \Bigg)^{-1} \notag\\
&\quad \cdot \Bigg( \sum_{i \in \V} \sizetilde \mY_{i,k}^{\Trans} \mR_{i,k}^{-1} (\vy_{i,k} - \sizehat \vLambda_{i,k}^* \mu_{i,k | k - 1}) \Bigg) \notag \\
&= \Phi \mW^{\frac{1}{2}} \Big( \left[\sizetilde \mY_{1,k}^{\Trans}, \dots, \sizetilde \mY_{\nS,k}^{\Trans} \right] \cdot \blkdiag\big(\mR_{1,k}^{-1}, \dots, \mR_{\nS,k}^{-1}\big) \notag \\
&\quad \cdot \left[\sizetilde \mY_{1,k}^{\Trans}, \dots, \sizetilde \mY_{\nS,k}^{\Trans} \right]^{\Trans} + \mI \Big)^{-1}
\left[\sizetilde \mY_{1,k}^{\Trans}, \dots, \sizetilde \mY_{\nS,k}^{\Trans} \right] \notag \\
&\quad \cdot \blkdiag\left(\mR_{1,k}^{-1}, \dots, \mR_{\nS,k}^{-1}\right) \Big(\left[\vy_{1,k}^{\Trans}, \dots, \vy_{\nS,k}^{\Trans}\right]^{\Trans} \notag \\
&\quad - \Big[\big(\sizehat \vLambda_{1,k}^* \mu_{1,k | k - 1}\big)^{\Trans}, \dots, \big(\sizehat \vLambda_{\nS,k}^* \mu_{\nS,k | k - 1}\big)^{\Trans}\Big]^{\Trans} \Big) \notag \\
&= \Phi \mW^{\frac{1}{2}} \big(\sizetilde \mY_k^{\Trans} \mR_k^{-1} \sizetilde \mY_k + \mI\big)^{-1} \sizetilde \mY_k^{\Trans} \mR_k^{-1} \big(\vy_k - \sizehat \vLambda_k^* \mu_{k | k - 1}\big) \notag \\
&= \Phi \mW^{\frac{1}{2}} \big(\mW^{\frac{1}{2}} \mY_k^{\Trans} \mR_k^{-1} \mY_k \mW^{\frac{1}{2}} + \mI\big)^{-1} \notag \\
&\quad \cdot \mW^{\frac{1}{2}} \mY_k^{\Trans} \mR_k^{-1} \big(\vy_k - \sizehat \vLambda_k^* \mu_{k | k - 1}\big) \notag \\
&= \mG_k \big(\vy_k - \sizehat \vLambda_k^* \mu_{k | k - 1}\big),
\end{align}
where $\mG_k = \Phi \mW^{\frac{1}{2}} \big(\mW^{\frac{1}{2}} \mY_k^{\Trans} \mR_k^{-1} \mY_k \mW^{\frac{1}{2}} + \mI\big)^{-1} \mW^{\frac{1}{2}} \mY_k^{\Trans} \mR_k^{-1}$. 
For any positive semi-definite matrix $\mP$, 
the following equality holds:
\begin{align}
(\mI + \mP)^{-1} = \mI - \mP (\mI + \mP)^{-1}.
\end{align}
Then, we have
\begin{align}
\mG_k
&= \Phi \mW^{\frac{1}{2}}  \Big( \mI - \mW^{\frac{1}{2}} \mY_k^{\Trans} \mR_k^{-1} \mY_k \mW^{\frac{1}{2}} \notag\\
&\quad \cdot \big(\mI + \mW^{\frac{1}{2}} \mY_k^{\Trans} \mR_k^{-1} \mY_k \mW^{\frac{1}{2}}\big)^{-1} \Big)
\mW^{\frac{1}{2}} \mY_k^{\Trans} \mR_k^{-1} \notag \\
&= \Phi \mW^{\frac{1}{2}}  \Big( \mW^{\frac{1}{2}} \mY_k^{\Trans} \mR_k^{-1} -  \mW^{\frac{1}{2}} \mY_k^{\Trans} \mR_k^{-1} \mY_k  \mW^{\frac{1}{2}} \notag \\
&\quad \cdot \big(\mI + \mW^{\frac{1}{2}} \mY_k^{\Trans} \mR_k^{-1} \mY_k \mW^{\frac{1}{2}}\big)^{-1} \mW^{\frac{1}{2}} \mY_k^{\Trans} \mR_k^{-1} \Big) \notag\\
&= \Phi \mW \mY_k^{\Trans} \Big( \mR_k^{-1} - \mR_k^{-1} \mY_k \mW^{\frac{1}{2}} \big(\mI + \mW^{\frac{1}{2}} \mY_k^{\Trans} \mR_k^{-1} \mY_k \mW^{\frac{1}{2}}\big)^{-1} \notag \\
&\quad \cdot \mW^{\frac{1}{2}} \mY_k^{\Trans} \mR_k^{-1} \Big).
\end{align}
Using Woodbury matrix identity (see Page 258 of \cite{Higham2002Accuracy}), from \eqref{eq:error covariance operator:dis} and \eqref{eq:error cov oper:cen}, we have
\begin{align}
\mG_k
&= \Phi \mW \mY_k^{\Trans} \big(\mY_k \mW \mY_k^{\Trans} + \mR_k\big)^{-1} \notag\\
&= \Phi \mW \Phi^* \Phi \gram^{-1} \mY_k^{\Trans} \big(\mY_k \gram^{-1} \Phi^* \Phi \mW \Phi^* \Phi \gram^{-1} \mY_k^{\Trans} + \mR_k\big)^{-1} \notag\\
&= \mP^{\phi}_{k | k - 1} \sizehat \vLambda_k (\sizehat \vLambda^*_k \mP^{\phi}_{k | k - 1} \sizehat \vLambda_k +  \mR_k)^{-1}.
\end{align}
This completes the proof.
\end{IEEEproof}

Theorem \ref{thm:converge} indicates that the proposed DNF-RKHS is capable of achieving the  the high accuracy of the centralized filtering while having the advantages of the distributed pattern.
Combining Proposition \ref{prop:KF} results in the following corollary.

\begin{corollary}
The DNF-RKHS converges to the centralized Kalman filter for linear dynamic system given by \eqref{eq:linear state space model}.
\end{corollary}

\begin{remark}
If $\{\vw_k\}$ and $\{\vv_{i,k}\}$ are mutually independent Gaussian white noise sequences, then the DNF-RKHS converges to the optimal filter in the sense of minimum mean squared error.
\end{remark}


\section{Examples}\label{sec:examples}

In this section, we provide two different target tracking scenarios to demonstrate the effectiveness of the proposed DNF-RKHS, compared with an effective DKF proposed in \cite{Chen2017}. 
We use the classical Gaussian and Laplace kernels given by \eqref{eq:Gaussian kernel} and \eqref{eq:Laplace kernel}, respectively, to estimate the KME of posterior distribution in the DNF-RKHS. For convenience, the DNF-RKHS equipped with the Gaussian and Laplace kernels are denoted as  DNF-RKHS(G) and DNF-RKHS(L), respectively. 

To measure the filters' estimation accuracy,
the root mean squared error (RMSE) and average Euclidean error (AEE) of state estimate are adopted, respectively. 
As analyzed convincingly in \cite{Rong2006Evaluation}, the AEE is generally better than the RMSE as a performance measure. The RMSEs and AEEs of position and velocity estimates are compared in the following two examples.
For a fair comparison purpose, all filtering algorithms were implemented by Octave on a computer with Intel Core i7 $2.60$ GHz processor. 

\begin{remark}
	We do not need to demonstrate the CNF-RKHS any more because the DNF-RKHS is iterated until convergence and thus has the same performance as the CNF-RKHS.
\end{remark}

\subsection{Target of Nearly Constant Velocity}

\textbf{1. Bearing-Only Measurement}

As in \cite{Ristic-Arulampalam-GordonGordon2004, Lan2015Nonlinear, Lan-Li-2017-Multiple}, we consider a target moving in a plane with a nearly constant velocity. The state transition equation is given by
\begin{align}\label{eq:TT:state transition}
	\vx_k = \mF_{k - 1} \vx_{k - 1} + \mG_{k - 1} \vw_{k - 1}, 
\end{align}
where $\vx_k = [x_k, \dot{x}_k, y_k, \dot{y}_k]^{\Trans}$, $[x_k, y_k]^{\Trans}$ and $[\dot{x}_k, \dot{y}_k]^{\Trans}$ are the position and velocity of the target, respectively, and
\begin{align}
	\mF_{k - 1} &= \blkdiag (\mF, \mF), \mG_{k - 1} = \blkdiag (\mG, \mG),\label{eq:Fk,Gk}\\
	\mF &=
	\begin{bmatrix}
		1 & \Delta t\\
		0 & 1
	\end{bmatrix},
	\mG =
	\begin{bmatrix}
	\Delta	t^2/2 \\
	\Delta	t
	\end{bmatrix},\label{eq:F,G}
\end{align}
with sampling period $\Delta t = 1\mathrm{s}$, the process noise $\vw_{k - 1}$ follows the Gaussian distribution $\mathcal{N}(0, \mQ_{k - 1})$ with $\mQ_{k - 1} = 10^2 G G^T \mathrm{(m^2/s^2)^2}$. The initial state was generated from the Gaussian distribution $\mathcal{N}(\bar \vx_0, P_0)$ with $\bar \vx_0 = [-500 \mathrm{m}, 18 \mathrm{m/s}, 500 \mathrm{m}, -12 \mathrm{m/s}]^{\Trans}$ and 
\begin{align}
P_0 = \diag([100 \mathrm{m^2}, 10 \mathrm{m^2/s^2}, 100 \mathrm{m^2}, 10 \mathrm{m^2/s^2}]^{\Trans}). \label{eq:TT_initial_P0}
\end{align}

The distributed sensor network consists of 10 sensors. For each node $i = 1,  2, \dots, 10$, the measurement at time $k$ includes an bearing:
\begin{align}
	\theta_{i,k} = \arctan \left(\frac{y_k - b_i}{x_k - a_i}\right) + v_{i,k},
\end{align}
where $[a_i\mathrm{m}, b_i\mathrm{m}]^{\Trans}$ is the position of Node $i$, and is generated randomly from the square area with center $[0\mathrm{m},0\mathrm{m}]^{\Trans}$ and side length $d = 5000\mathrm{m}$, the measurement noise $v_{i,k}$ is distributed from the Gaussian distribution $\N(0, \sigma_{i,k})$ with $\sigma_{i,k} = 0.01 \mathrm{rad^2}$. We take $\sigma = 1$ in the Gaussian and Laplace kernels and $50$ sample points in the DNF-RKHS. 

\begin{figure}[htbp]
	\centering
	\includegraphics[width=3.7in] {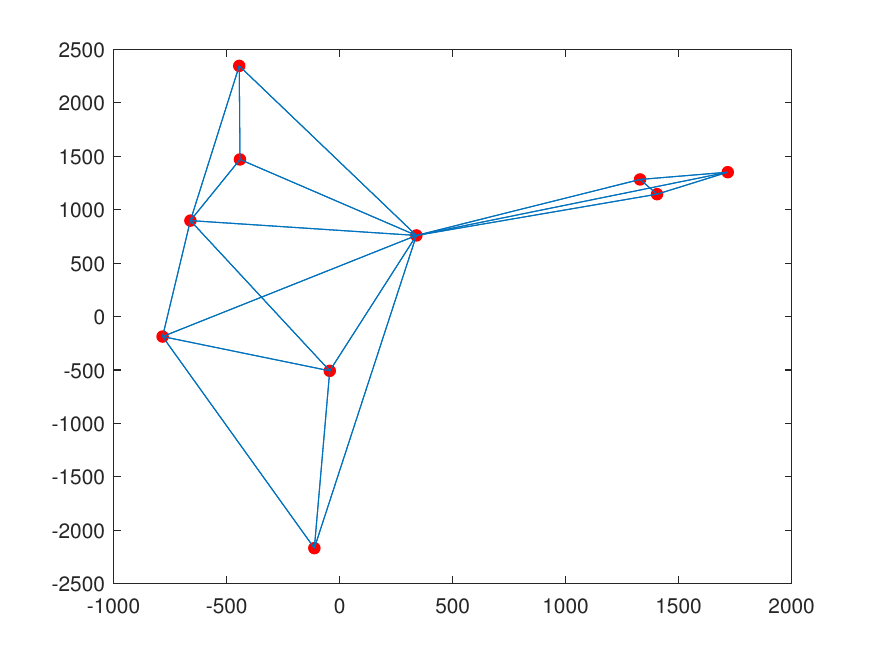}
	\caption{Communication Topology of Sensor Network (Example A.1).}
	\label{fig:TT_topology}
\end{figure}

\begin{figure}[htbp]
	\centering
	\includegraphics[width=3.7in] {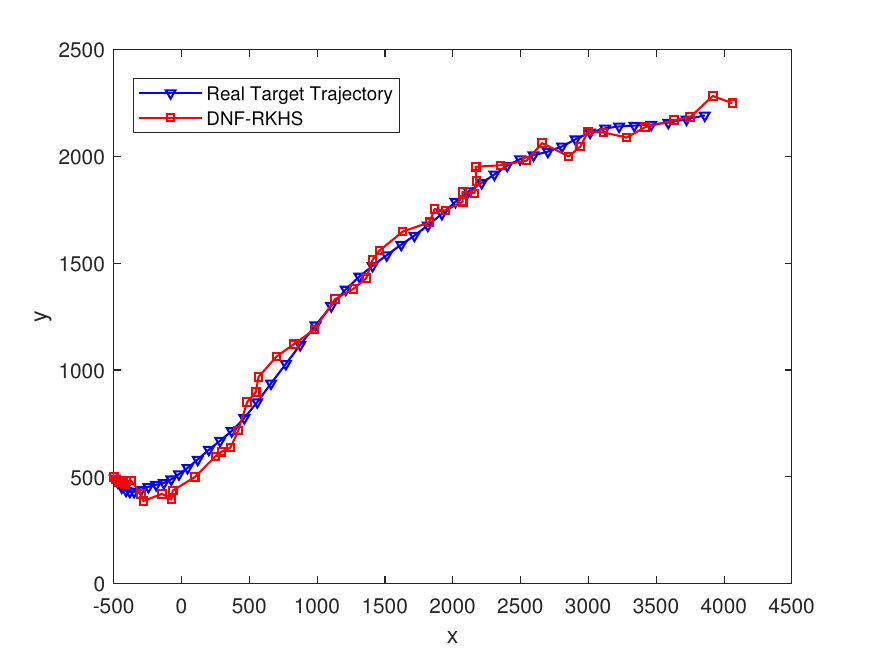}
	\caption{Real Target Trajectory and Estimate by DNF-RKHS (Example A.1).}
	\label{fig:TT_StateEstimate}
\end{figure}

Fig.~\ref{fig:TT_topology} shows the communication topology of the sensor network, where the red dots represent the sensors and the blue lines are the links among them. It is not difficult to find that Fig.~\ref{fig:TT_topology} satisfies Assumption  \ref{assum:topology}.
Fig.~\ref{fig:TT_StateEstimate} depicts the real trajectory of the target and the estimate by the DNF-RKHS over 50 time steps. It is clearly observed that the DNF-RKHS shows good estimation performance. 

Table \ref{tab:TT_Comparison} shows the averaged RMSEs and AEEs of position and velocity estimates over 100 Monte Carlo runs and 50 time steps. 
Compared with the DKF, the DNF-RKHS(G) and DNF-RKHS(L) have respectively improved by $-5\%$ and $38\%$ in RMSE of position estimate, by $49\%$ and $26\%$ in RMSE of velocity estimate, by $10\%$ and $26\%$ in AEE of position estimate, and by $41\%$ and $12\%$ in AEE of velocity estimate. 
In result, the DNF-RKHS(G) and DNF-RKHS(L)  achieve more accurate position and velocity estimates than the DKF in both RMSE and AEE sense. More specifically, the DNF-RKHS(G) and  DNF-RKHS(L) have the best velocity and position estimates, respectively.
Interestingly, the DNF-RKHS(G) has a relatively large RMSE of position estimate (second row, third column) among the three compared filters. This could be caused by an estimate with a relatively large deviation at a certain time since the RMSE is susceptible to an individual point with large deviation.  Meanwhile, the DNF-RKHS(G) has smaller AEE of position estimate (fourth row, third column) than the DKF. Then, we can infer that the DNF-RKHS(G) produces good state estimates in most of the time except for some single moment.

In fact, the numerical results in Table \ref{tab:TT_Comparison} have corresponding theoretical supports: while the DKF derives moment estimates of the posterior distribution, the DNF-RKHS provides the posterior density estimate and thus results in a significantly improved filtering performance. 

\begin{table}[htbp]
	\renewcommand{\arraystretch}{1.5}
	\caption{Comparison Results of Averaged RMSEs and AEEs of Position and Velocity Estimates (Example A.1).}
	\setlength{\tabcolsep}{0.2mm}
	\label{tab:TT_Comparison}
	\centering
	\begin{tabular}{c|c|c|c}
		\hline
		\diagbox{Measure}{Error}{Method} & DKF & DNF-RKHS(G) & DNF-RKHS(L)  \\
		\hline
		RMSEs of Position Estimates ($\mathrm{m}$)   & $702.14$ & $739.93$ &  $435.21$  \\
		RMSEs of Velocity Estimates ($\mathrm{m/s}$) & $118.22$ & $60.24$  &  $87.91$  \\
	    AEEs of Position Estimates ($\mathrm{m}$)    & $324.57$ & $291.70$ &  $240.92$  \\
	    AEEs of Velocity Estimates ($\mathrm{m/s}$)  & $63.88$  & $37.92$  &  $56.32$  \\
		\hline
	\end{tabular}
\end{table}

\textbf{2. Range and Bearing Measurements}

The motion of the target is also described by \eqref{eq:TT:state transition}--\eqref{eq:F,G} with sampling period $\Delta t = 0.08\mathrm{s}$. Besides, the initial state was drawn from Gaussian distribution $\N(\bar \vx_0, \mP_0)$ with  $\bar \vx_0 = [0 \mathrm{m}, -18 \mathrm{m/s}, 500 \mathrm{m}, 12 \mathrm{m/s}]^{\Trans}$ and $P_0$ given by \eqref{eq:TT_initial_P0}.

\begin{figure}[htbp]
	\centering
	\includegraphics[width=3.7in] {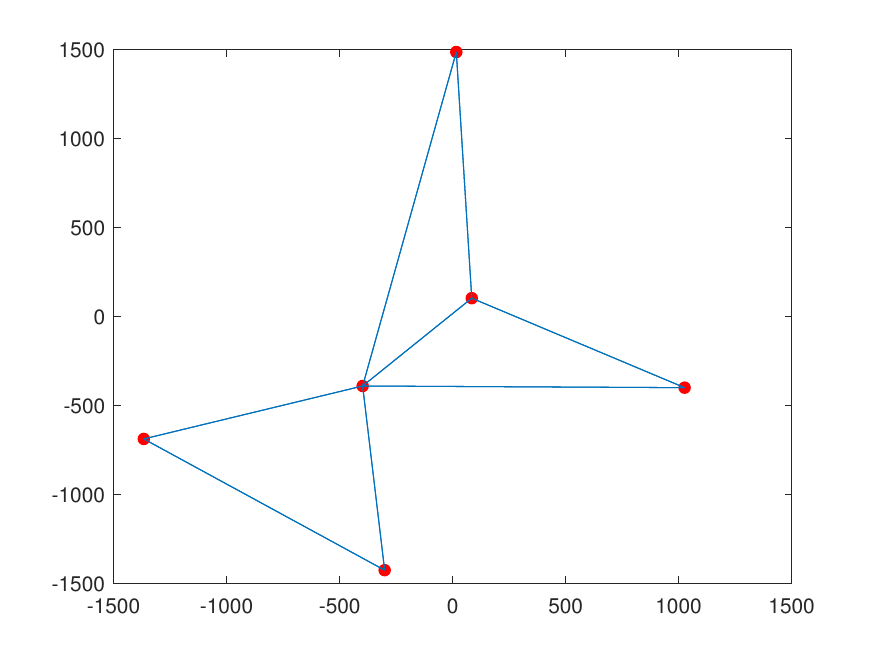}
	\caption{Communication Topology of Sensor Network (Example A.2).}
	\label{fig:PCO_topology}
\end{figure}

As in Fig.~\ref{fig:PCO_topology}, we consider a sensor network consisting of 6 sensors, where the $i$-th node ($i = 1, 2, \dots, 6$), located at $[a_i\mathrm{m}, b_i\mathrm{m}]^{\Trans}$, is randomly distributed in the square area of length $d = 3000\mathrm{m}$ centered on $[0\mathrm{m}, 0\mathrm{m}]^{\Trans}$. At time $k$, Node $i$ observes the range $r_{i,k}$ and bearing $\theta_{i,k}$ of the target, and the measurement equation is 
\begin{align}
\begin{bmatrix}
	r_{i,k} \\
	\theta_{i,k}
\end{bmatrix}
&=
\begin{bmatrix}
	\sqrt{(x_k - a_i)^2 + (y_k - b_i)^2}\\
	\arctan\big(\frac{y_k - b_i}{x_k - a_i}\big)
\end{bmatrix}
+ \vv_{i,k},
\end{align}
where the measurement noise $\{\vv_{i,k}\}$ is the zero-mean Gaussian white noise sequence with covariance matrix $\mR_{i,k} = \diag([100\mathrm{m^2}, 0.01\mathrm{rad^2}]^{\Trans})$. We take $\sigma = 2$ in the Gaussian and Laplace kernels and $50$ sample points in the DNF-RKHS.

\begin{figure}[htbp]
	\centering
	\includegraphics[width=3.7in] {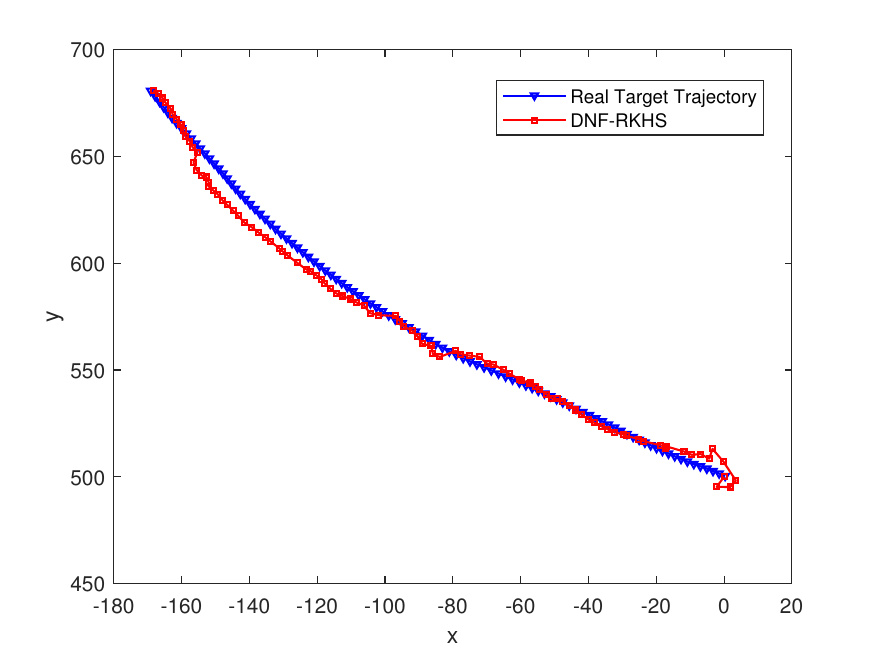}
	\caption{Real Target Trajectory and Estimate by DNF-RKHS (Example A.2).}
	\label{fig:PCO_StateEstimate}
\end{figure}

The tracking performance of the DNF-RKHS is shown in 
Fig.~\ref{fig:PCO_StateEstimate}, where the target moving at a nearly constant velocity for 100 time steps can be clearly seen. 
Overall, the DNF-RKHS shows a satisfactory filtering performance, except for some small yet acceptable deviations.
The comparison results of position and velocity estimates over 100 time steps and 100 Monte Carlo runs in the averaged RMSE and AEE sense are given in Table~\ref{tab:PCO_Comparison}. 
Compared with the DKF, the DNF-RKHS(G) and DNF-RKHS(L) have respectively improved by $34\%$ and $37\%$ in RMSE of position estimate, by $35\%$ and $35\%$ in RMSE of velocity estimate, by $1\%$ and $10\%$ in AEE of position estimate, and by $12\%$ and $10\%$ in AEE of velocity estimate. 
Like Example A.1, the DNF-RKHS(G) and DNF-RKHS(L) both have higher filtering accuracy than the DKF, but they show a broadly similar performance in this example. This indicates that the choice of kernel function may not be the decisive factor affecting the filtering performance of the DNF-RKHS in this particular scenario.

\begin{table}[htbp]
	\renewcommand{\arraystretch}{1.5}
	\caption{Comparison Results of Averaged RMSEs and AEEs of Position and Velocity Estimates (Example A.2).}
	\setlength{\tabcolsep}{0.3mm}
	\label{tab:PCO_Comparison}
	\centering
	\begin{tabular}{c|c|c|c}
		\hline
		\diagbox{Measure}{Error}{Method} & DKF & DNF-RKHS(G) & DNF-RKHS(L)  \\
		\hline
		RMSEs of Position Estimates ($\mathrm{m}$)   & $ 6.80$  & $4.49$   &  $4.26$  \\
		RMSEs of Velocity Estimates ($\mathrm{m/s}$) & $8.25$   & $5.34$   &  $5.37$  \\
		AEEs of Position Estimates ($\mathrm{m}$)    & $4.03$   & $3.98$   &  $3.63$  \\
		AEEs of Velocity Estimates ($\mathrm{m/s}$)  & $5.45$   & $4.82$   &  $4.88$  \\
		\hline
	\end{tabular}
\end{table}

\subsection{Aircraft Coordinated Turn}

As in \cite{JIA2013High}, we address the following dynamic system:
\begin{align*}
	\vx_k &=
	\begin{bmatrix}
		1 & \frac{\sin(\omega_{k - 1} \Delta t)}{\omega_{k - 1}} & 0 & \frac{\cos(\omega_{k - 1} \Delta t) - 1}{\omega_{k - 1}} & 0 \\
		0 & \cos(\omega_{k - 1} \Delta t) & 0 & - \sin(\omega_{k - 1} \Delta t) & 0 \\
		0 & \frac{1 - \cos(\omega_{k - 1} \Delta t)}{\omega_{k - 1}} & 1 & \frac{\sin(\omega_{k - 1} \Delta t)}{\omega_{k - 1}} & 0 \\
		0 & \sin(\omega_{k - 1} \Delta t) & 0 & \cos(\omega_{k - 1} \Delta t) & 0 \\
		0 & 0 & 0 & 0 & 1 
	\end{bmatrix}
	\vx_{k - 1} \notag\\
	&\quad + \vw_{k - 1},
\end{align*}
where $\omega_{k - 1}$ is an unknown turn rate at time $k - 1$, $\Delta t = 0.2$ is sampling period, and $\{\vw_{k}\}$ is the Gaussian white noise sequence with zero mean and covariance matrix
\begin{align*}
	\mQ_{k - 1} = q
	\begin{bmatrix}
		\frac{\Delta t^3}{3} & \frac{\Delta t^2}{2} & 0 & 0 & 0 \\
		\frac{\Delta t^2}{2} & \Delta t & 0 & 0 & 0 \\
		0 & 0 & \frac{\Delta t^3}{3} & \frac{\Delta t^2}{2} & 0 \\
		0 & 0 & \frac{\Delta t^2}{2} & \Delta t & 0 \\
		0 & 0 & 0 & 0 & 1.75 \times 10^{-3} \Delta t
	\end{bmatrix},
\end{align*}
where $q$ is a scalar with respect to noise intensity. The initial state of the aircraft was distributed from $\N(\bar \vx_0, \mP_0)$ with $\bar \vx_0 = [5000\mathrm{m}, 180\mathrm{m/s}, 5000\mathrm{m}, 180\mathrm{m/s}, 0.01\mathrm{rad}]$ and $\mP_0 = \diag([1000\mathrm{m^2}, 100\mathrm{m^2 /s^2}, 1000\mathrm{m^2}, 100\mathrm{m^2 /s^2}, 0.001\mathrm{rad^2}]^{\Trans})$.

\begin{figure}[htbp]
	\centering
	\includegraphics[width=3.7in] {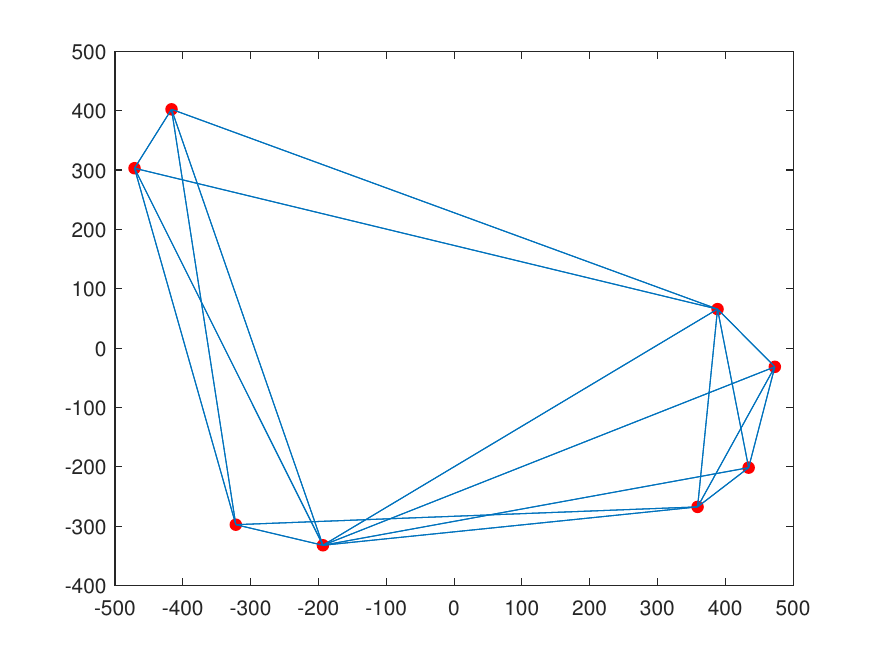}
	\caption{Communication Topology of Sensor Network (Example B).}
	\label{fig:ACT_Topology}
\end{figure}

The measurements of range $r_{i,k}$, bearing $\theta_{i,k}$ and range-rate $\dot r_{i,k}$ are taken by 8 sensors shown in Fig.~\ref{fig:ACT_Topology}, where the side length $d = 1000\mathrm{m}$. For each node $i = 1, \dots, 8$, the measurement equation is 
\begin{align}
	\begin{bmatrix}
		r_{i,k} \\
		\theta_{i,k} \\
		\dot r_{i,k}
	\end{bmatrix}
	&=
	\begin{bmatrix}
		\sqrt{(x_k - a_i)^2 + (y_k - b_i)^2}\\
		\arctan\big(\frac{y_k - b_i}{x_k - a_i}\big) \\
		\frac{(x_k - a_i) \dot x_k + (y_k - b_i) \dot y_k}{\sqrt{(x_k - a_i)^2 + (y_k - b_i)^2}}
	\end{bmatrix}
	+ \vv_{i,k},
\end{align}
with $\vv_{i,k} \sim \N(0, \diag([10000\mathrm{m^2},0.01\mathrm{rad^2},10\mathrm{m^2 /s^2}]^{\Trans}))$. 

\begin{figure}[htbp]
	\centering
	\includegraphics[width=3.7in] {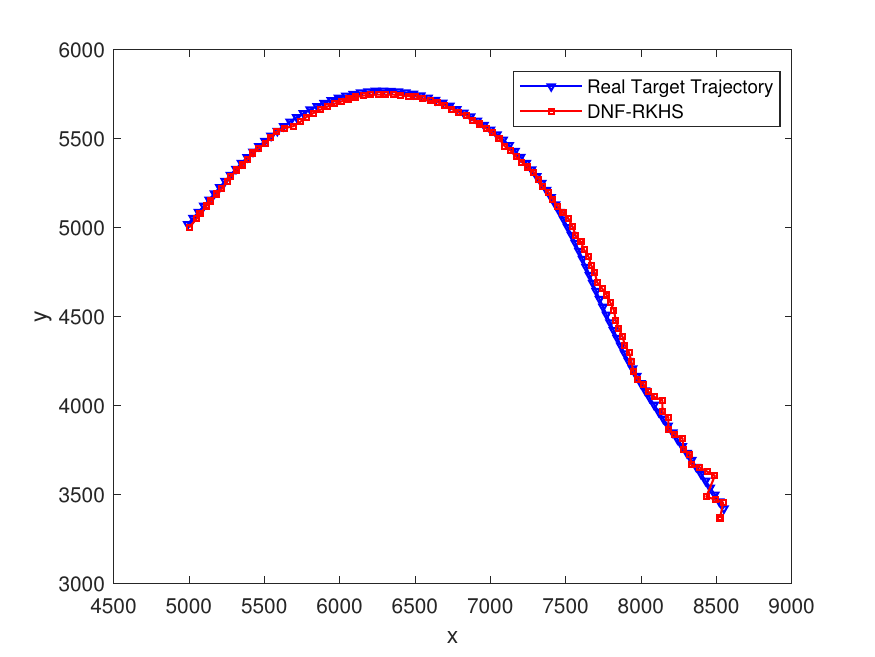}
	\caption{Real Target Trajectory and Estimate by DNF-RKHS (Example B).}
	\label{fig:ACT_StateEstimate}
\end{figure}

By taking $q =1\mathrm{m^2 /s^3}$, $\sigma = 0.01$ in the Gaussian and Laplace kernels, and $60$ sample points in the DNF-RKHS, the filtering performance of the DNF-RKHS is given in Fig.~\ref{fig:ACT_StateEstimate}. As can be seen, the DNF-RKHS shows pretty good tracking performance in this example. The averaged RMSEs and AEEs of position and velocity estimates averaging over 100 time steps and 50 Monte Carlo runs are given in Table~\ref{tab:ACT_Comparison}. Unlike Examples A.1 and A.2, the DNF-RKHS(L) shows the best performance among the three compared filters in both position and velocity estimation. Compared with the DKF, the DNF-RKHS(G) and DNF-RKHS(L) have respectively improved by $15\%$ and $24\%$ in RMSE of position estimate, by $30\%$ and $37\%$ in RMSE of velocity estimate, by $18\%$ and $23\%$ in AEE of position estimate, and by $29\%$ and $34\%$ in AEE of velocity estimate. This suggests a large performance improvement by the DNF-RKHS(G) and DNF-RKHS(L) over the DKF, especially by the DNF-RKHS(L).

\begin{table}[htbp]
	\renewcommand{\arraystretch}{1.5}
	\caption{Comparison Results of Averaged RMSEs and AEEs of Position and Velocity Estimates (Example B).}
	\setlength{\tabcolsep}{0.2mm}
	\label{tab:ACT_Comparison}
	\centering
	\begin{tabular}{c|c|c|c}
		\hline
		\diagbox{Measure}{Error}{Method} & DKF & DNF-RKHS(G) & DNF-RKHS(L)  \\
		\hline
		RMSEs of Position Estimates ($\mathrm{m}$)   & $ 89.83$  & $75.96$   &  $ 67.90$  \\
		RMSEs of Velocity Estimates ($\mathrm{m/s}$) & $ 58.40$   & $40.85$   &  $36.99$  \\
		AEEs of Position Estimates ($\mathrm{m}$)    & $80.48$   & $66.28$   &  $61.99$  \\
		AEEs of Velocity Estimates ($\mathrm{m/s}$)  & $ 50.08$   & $35.57$   &  $32.82$  \\
		\hline
	\end{tabular}
\end{table}

\section{Conclusion}\label{sec:conclusion}

In this paper, we have proposed the DNF-RKHS by approximating the posterior density with KME for distributed nonlinear dynamic systems. 
By introducing the higher-dimensional RKHS, the nonlinear measurement function has been linearly converted, and the update rule of KME has been well established in the RKHS.
To show the high estimation accuracy of the DNF-RKHS, we have developed the CNF-RKHS and further proved that the DNF-RKHS converges to the CNF-RKHS. 
By adopting KME, the DNF-RKHS achieves the centralized estimation accuracy while ensuring the distributed pattern.
Finally, two different target tracking examples have demonstrated the effectiveness of the DNF-RKHS in different scenarios.

For distributed nonlinear dynamic systems, the issue of privacy may occur due to the information interaction between sensors, so the privacy-preserving distributed nonlinear filtering will be studied in the future.

\bibliographystyle{IEEEtran}
\bibliography{bibfile}


%

%
%
%
%
%

\ifCLASSOPTIONcaptionsoff
  \newpage
\fi

\end{document}